\documentclass[conference]{IEEEtran}
\IEEEoverridecommandlockouts
\usepackage{cite}
\usepackage{amsmath,amssymb,amsfonts}
\usepackage{algorithmic}
\usepackage{graphicx}
\usepackage{textcomp}
\usepackage{xcolor}
\def\BibTeX{{\rm B\kern-.05em{\sc i\kern-.025em b}\kern-.08em
    T\kern-.1667em\lower.7ex\hbox{E}\kern-.125emX}}

\usepackage{fancyhdr}
\begin{document}

\title{Federated Deep Reinforcement Learning for Resource Allocation in O-RAN Slicing \\
{
}
}
\author{\IEEEauthorblockN{Han Zhang, Hao Zhou, and Melike Erol-Kantarci, \IEEEmembership{Senior Member, IEEE}}
\IEEEauthorblockA{\textit{School of Electrical Engineering and Computer Science,}
\textit{University of Ottawa}\\
Emails:\{hzhan363, hzhou098, melike.erolkantarci\}@uottawa.ca}}

\maketitle
\thispagestyle{fancy}   
\fancyhead{}                
\lhead{Accepted by 2022 IEEE Global Communications Conference (GLOBECOM), \copyright2022 IEEE}
\cfoot{}
\renewcommand{\headrulewidth}{0pt}      

\begin{abstract}
Recently, open radio access network (O-RAN) has become a promising technology to provide an open environment for network vendors and operators. Coordinating the x-applications (xAPPs) is critical to increase flexibility and guarantee high overall network performance in O-RAN. Meanwhile, federated reinforcement learning has been proposed as a promising technique to enhance the collaboration among distributed reinforcement learning agents and improve learning efficiency. In this paper, we propose a federated deep reinforcement learning algorithm to coordinate multiple independent xAPPs in O-RAN for network slicing. We design two xAPPs, namely a power control xAPP and a slice-based resource allocation xAPP, and we use a federated learning model to coordinate two xAPP agents to enhance learning efficiency and improve network performance. Compared with conventional deep reinforcement learning, our proposed algorithm can achieve 11\% higher throughput for enhanced mobile broadband (eMBB) slices and 33\% lower delay for ultra-reliable low-latency communication (URLLC) slices.
\end{abstract}


\section{Introduction} 

One of the main purposes of Open radio access network (O-RAN) is to improve RAN performance and supply chain by adopting open interfaces that allow multiple vendors to manage the intelligence of RAN. O-RAN is expected to support interoperability between devices from multiple vendors and provide network flexibility at a lower cost \cite{b1-a}. On the other hand, network slicing has been considered as a promising technique that can separate a physical network into multiple logical networks to provide service-customized solutions \cite{b1}. It can be used for multi-service problems in O-RAN to maintain slice-level key performance indicators. 



In O-RAN, there are various  network functions called as x-applications (xAPPs). Each xAPP can be considered as an independent agent to control a specific network function such as power control and resource allocation \cite{b1-c}. Considering that these xAPPs can be designed by various vendors, xAPP can apply conflicting configurations when performing independent optimization tasks, thus leading to overall network performance degradation\cite{b1-b} . Moreover, in the context of network slicing, each xAPP will serve multiple network slices with diverse quality of service (QoS) requirements, which will significantly increase the complexity of network management and control. Therefore, coordinating various xAPP agents and satisfying the service level agreements of multiple slices simultaneously is a crucial challenge for O-RAN.

Machine learning, especially reinforcement learning (RL), has been widely used to solve optimization problems in wireless networks\cite{b3}. In RL, individual agents interact with their environment to gain experience and learn to get the maximum expected reward. 
Meanwhile, federated learning (FL) is an emerging approach in machine learning, which refers to multiple partners training models separately in a decentralized but collaborative manner to build shared models while maintaining data privacy \cite{b4}. Federated reinforcement learning is proposed as a combination of FL and RL, aiming to apply the idea of FL to distributed RL models to enhance the collaboration among multiple RL agents \cite{b5}. Considering the openness and intelligence requirements of the O-RAN environment, federated reinforcement learning becomes an interesting solution to coordinate the operation of multiple intelligent xAPP agents in O-RAN.

Currently, there have been some applications of federated reinforcement learning in wireless communications area. However, in most existing works, different agents are designed to perform similar network applications with the same action and state spaces to accelerate the exploration of the environment through experience sharing \cite{b11}-\cite{b13}. To the best of our knowledge, there are limited works on federated reinforcement learning on the problems where multiple agents have different action and state spaces and perform different network applications.

In this work, we propose a federated deep reinforcement learning algorithm to coordinate multiple xAPPs for network slicing in O-RAN . We first define two xAPPs: a power control xAPP and a hierarchical radio resource allocation xAPP. The local models are trained at each xAPP agent, and then these local models are submitted to a coordination model and serve as the input for predicting a joint Q-table. Finally, the joint Q-table is disassembled into local Q-tables for the actions. Simulation results show that our proposed federated deep reinforcement learning model can obtain 11\% higher throughput and 33\% lower delay than independent learning models.

The rest of the paper is arranged as follows. In Section \ref{s2}, related works are introduced, and the system model is described in Section \ref{s3}. Section \ref{s4} explains the implementation of two xAPP applications. Section \ref{s5} introduces the proposed federated deep reinforcement learning algorithm. Simulation settings and results are introduced in Section \ref{s6}, and Section \ref{s7} concludes the paper.

\section{Related Works}
\label{s2}
There have been many studies that applies RL techniques to power control and resource allocation problems in 5G networks. In \cite{b7}, a deep Q-learning algorithm is performed for the combined optimization of power control, beam forming, and interference reduction. In \cite{b9}, the communication and computational resources are distributed to randomly arriving slice requests with different weights and QoS requirements with a Q-learning algorithm. However, in these works, multiple network applications are generally combined into one single agent for joint optimization. They are not applicable to cases when different applications are considered as various independent agents in the O-RAN environment. 

On the other hand, some existing works studied possible applications of federated reinforcement learning in the area of wireless communication. In \cite{b11}, the authors leveraged deep reinforcement learning to allocate heterogeneous resources and perform computation offloading in the 5G ultra-dense network scenarios and the RL model is trained in a decentralized way with FL. In \cite{b12}, a hybrid federated reinforcement learning-based structure is proposed for devices association in RAN slicing with parameter aggregation on two levels. In \cite{b13}, a federated reinforcement learning algorithm is used to control the access of users in O-RAN, where RL is used for access decision making on single users and FL is used to aggregate models from different users. In these works, 
FL is mainly adopted as a solution for the federation between agents performing the same network functions, which often have similar state and action spaces, without considering how different network functions and agents are federated.

In our previous work \cite{b14}, an information exchanging-based team learning algorithm is proposed to mitigate the conflicts between xAPPs. However, in this work, we include multiple network slices with diverse QoS requirements and apply a novel federated deep reinforcement learning algorithm to enhance network performance.


\section{System Model}
\label{s3}
\begin{figure}[!t]
\centerline{\includegraphics[width=3.0in]{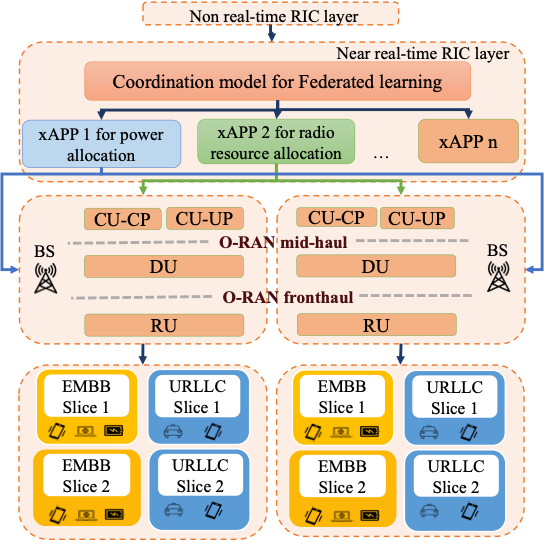}}
\caption{O-RAN network system model.}
\label{fig1}
\vspace{-15pt}
\end{figure}

As shown in Fig.~\ref{fig1}, we consider an O-RAN based deployment that includes multiple network functions and multiple slices for downlink transmission. The O-RAN system is composed of two radio intelligent controllers (RIC), non real-time RIC and near real-time RIC, centralized unit (CU), distributed unit (DU) and radio unit (RU). The CU is further decoupled into control plane (CP) and user plane (UP). Each base station (BS) includes two types of slices, namely eMBB slice, and URLLC slice, and each slice may contain several UEs with similar QoS requirements. On the other hand, two network functions are jointly considered: power control and radio resource allocation. We aim to satisfy the diverse QoS demand of slices by jointly controlling these network functions. The proposed scheme can be applied for any other xAPP-based network function coordination. But here, we consider power control and resource allocation since they are among the most fundamental network control functions.


For the power control xAPP, it decides the transmission power level of each BS to control its own channel capacity and the interference on adjacent BSs.
For the resource allocation xAPP, we consider the radio resource block (RB) as the smallest resource unit to be allocated \cite{b15}. 
The RBs are first allocated to each slice, then the intra-slice RB allocation is implemented to distribute RBs to specific user device transmissions. On the upper level, a coordination model in near real-time RIC is set up to federate two xAPPs. The goal of the model is to coordinate two xAPPs to minimize the delay of URLLC slices and to maximize the throughput for eMBB slices. 

We assume that at time slot $t$, the transmission power of the $r^{th}$ RB of BS $k$ is denoted by $P_{k,r}$. The signal interference noise ratio (SINR) of the transmission link between BS $k$ and device $m$ on the $r^{th}$ RB can be formulated as:
\begin{equation}
\eta_{k,r,m}=\frac{\alpha_{k,r,m} g_{k,m} P_{k,r}}{\sum_{k'\in K, k'\neq k}\sum_{m'\in m_{k'}} \alpha_{k',r,m'} g_{k',m} P_{k',r}+B_{r}N_{0}},\label{eq1}
\end{equation}  
where $\alpha_{k,r,m}$ is a binary indicator that denotes whether the $r^{th}$ RB of BS $k$ is allocated to device $m$. $g_{k,m}$ is the channel gain between BS $k$ and device $m$. $B_{r}$ denotes the bandwidth of the $r^{th}$ RB and $N_{0}$ denotes the noise power density.

The link capacity between BS $k$ and device $m$ is denoted as $C_{k,m}$ and it can be calculated as:

\begin{equation}
C_{k,m} = \sum_{r\in R} B_{r} log_{2}(1+\eta_{k,r,m}),\label{eq2}
\end{equation}
where $R$ denotes the set of all the RBs that can be allocated.

The total delay of user $m$ consists of three components, the transmission delay, the queuing delay and the retransmission delay. It can be given as:
\begin{equation}
d^{total}_{m} = d^{tx}_{m} + d^{que}_{m} + d^{rtx}_{m},\label{eq3}
\end{equation}
where $d^{tx}_{m}$, $d^{que}_{m}$ and $d^{rtx}_{m}$ denote the transmission delay, the queuing delay and the retransmission delay. The transmission delay is formulated as:
\begin{equation}
d^{tx}_{m} = \frac{L_{m}}{C_{k,m}},\label{eq4}
\end{equation}
where $L_{m}$ is the packet size of user $m$. For eMBB slices, optimization aims to achieve maximum total throughput, and for URLLC slices, the aim is to minimize the average delay of packets.

The problem can be formulated as follows:
\begin{align}
 \underset{P_{k},\alpha_{k,r,m}}{max}\ &\sum_{k\in K}\sum_{n\in N_{k}}w_{n}r_{n} \label{eq5}\\
s.t.\ &(\ref{eq1})-(\ref{eq4})\nonumber
\\& r_n^{embb} = \begin{cases}
tan^{-1}(\underset{m\in M_{n}^{embb}}{\sum}b_m),&|H_{n}^{embb}|\neq0\\
0,&else \\
\end{cases}\tag{5a}\label{eq5a}
\\& r_n^{urllc} = \begin{cases}
\underset{m\in M_{n}^{urllc}}{1-\sum}d_m,&|H_{n}^{urllc}|\neq0\\
0,&else\\
\end{cases}\tag{5b}\label{eq5b}
\\& P_{min} \leq P_{k} \leq P_{max}, \forall k\tag{5c}\label{eq5c}
\\& \alpha_{k,r,m} = \{0,1\}, \forall k, r, m\tag{5d}\label{eq5d}
\\& \Sigma_{m \in M}\alpha_{k,r,m} = 1, \forall k,r\tag{5e}\label{eq5e}
\end{align}
where $w_n$ and $r_n$ denotes the priority weight and the reward of each slice respectively. Equation \eqref{eq5a} shows that the reward of eMBB slices is decided by the total throughput. $M_{n}^{embb}$ denotes the set of devices in the $n^{th}$ slice and $H_{n}^{urllc}$ denotes the queued length of packets in slice $n^{th}$. The inverse tangent calculator is used to mapping the value to a limited interval. Equation \eqref{eq5b} shows that the reward of URLLC slices is decided by the packet delay. Equation \eqref{eq5c}, \eqref{eq5d} and \eqref{eq5e} ensure the assigned power level and allocated RBs are limited by the maximum power constraints and the resource constraints.

\section{DQN based Power Control and Resource Allocation xAPPs}
\label{s4}
In this section, we introduce how to transform the power control and radio resource allocation as two independent xAPPs. For power control, we deploy a deep Q-Networks (DQN) to decide the transmission power level of the BS. For radio resource allocation, we combine the DQN with a hierarchical decision-making mechanism for RBs allocation. 

\subsection{Power Control xAPP Agent}
The power control xAPP will decide the transmission power level of the BS, and after that we assume the power is uniformly allocated to all RBs. The Markov decision process (MDP) of power control agent is defined as follows:
\begin{itemize}
    \item State: The state of power control model includes the queue length of packets, the current delay and the current transmission power, which is given as:
    \begin{equation}
    s_{k,t} = \{H_{n},\sum_{m\in M_{n}}d_{m}, P_{k}|\forall n \in N, k \in K\},\label{eq6}
    \end{equation}
    where $H_{n}$ denotes the queue length of packets in the $n^{th}$ slice, $P_{k}$ denotes the transmission power of the $k^{th}$ BS. Here the state definition represents the traffic demand and network status, and then the agent can change the transmission power accordingly to achieve desired network performance.
    \item Action: The action of power control is to choose power level from:
    \begin{equation}
    a_{k,t} = \{1,2,...,L_{max}\},\label{eq7}
    \end{equation}
    where $L_{max}$ denotes the highest power level. Then the transmission power of the $k^{th}$ BS is given as:
    \begin{equation}
    P_{k,t} = \frac{a_{k,t}P_{max}}{L_{max}},\label{eq8}
    \end{equation}
    \item Reward: The reward of power control model is defined as the weighted sum reward of all the slices with a penalty of large transmission power.
    \begin{equation}
    r_{k,t} = \sum_{n\in N_{k}}w_{n}r_{n}-\alpha a_{k,t}, \label{eq9}
    \end{equation}
    The purpose of establishing penalty items is to
    balance the power consumption and network performance, and control the interference on neighbouring BSs.

\end{itemize}

The goal of DQN is to maximize the expected long-term reward, which is given as:
\begin{equation}
Q(s,a) = E[r_t + \gamma Q(s_{t+1},a_{t+1})|s_t = s, a_t = a],\label{eq10}
\end{equation}

In DQN, the Q-values are approximated by a deep neural network (DNN), and the stochastic gradient descent algorithm is adopted to update the parameters of DNN, which can be given as:
\begin{equation}
\begin{split}
\theta_{t+1} &= \theta_t + \alpha[r_t + \gamma \mathop{max}\limits_{a'}Q(s_{t+1},a_{t+1};\theta_t) \\ & -Q(s_t,a_t;\theta_t)]\nabla Q(s_t,a_t;\theta_t),\label{eq11}
\end{split}
\end{equation}
where $\theta$,$\alpha$ and $\gamma$ respectively denote the parameters of DQN, the learning rate and the discount factor. By experience replay and updating the parameters iteratively, the DNN can predict the expected accumulated reward and choose the optimal action accordingly.

\subsection{Radio Resource Allocation xAPP Agent}
For radio resource allocation xAPP, the inter-slice radio resource allocation is first performed to decide the amount of RBs that will be allocated to each slice. The MDP of inter-slice radio resource allocation is given as follows:

\begin{itemize}
    \item State: The state is composed of the queue length of packets and the current delay, which is given as:
    \begin{equation}
    s_{k,t} = \{H_{n},\sum_{m\in M_{n}}d_{m}|\forall n \in N\},\label{eq6-I}
    \end{equation}
    \item Action: The action is to decide the portion of RBs allocated to each slice, which is given as:
    \begin{equation}
    a_{k,t} = \{RB_{n}\in\{0,1,...,R_{max}\}|\sum_{n \in N}RB_{n}=R_{max} \},\label{eq12}
    \end{equation}
    where $R_{max}$ denotes the number of RB groups that can be allocated
    \item Reward: The reward of inter-slice radio resource allocation model is defined as the weighted sum reward of all the slices, which is given as:
    \begin{equation}
    r_{k,t} = \sum_{n\in N_{k}}w_{n}r_{n}, \label{eq14}
    \end{equation}
\end{itemize}

Then, for the intra-slice resource allocation,  
we assume all the devices in the same slice have equal priority. Hence, we deploy a proportional fair algorithm (PPF)\cite{b21}.
The basic idea is to determine the priority of devices for resource scheduling based on the ratio of instantaneous transmission rate and the long-term average transmission rate of a single device.
The intra-slice radio resource allocation strategy based on proportional fairness scheduling can be formulated as follows:
\begin{align}
 \underset{m \in M_{k,n}}{argmax}\ & \frac{C_{k,m}}{(\sum^{t}_{t-\Delta t}b_{t,m})/\Delta t}
 \label{eq15}
\end{align}
where $(\sum^{t}_{t-\Delta t}b_{t,m})/\Delta t$ denotes the average transmission rate of device $m$ over the past period $\Delta t$. The RBs allocated to slice $n$ will be given to device $m$ according to the selection rule in the above formula.

\section{Federated Deep Reinforcement Learning for xAPP Coordination}
\label{s5}
In this section, we firstly introduce federated deep reinforcement learning, then illustrate how we deploy federated deep reinforcement learning for coordination of xAPPs. 

\subsection{Federated Deep Reinforcement Learning}

Federated deep reinforcement learning is proposed as a combination of RL and FL. In federated deep reinforcement learning, the samples, features, and labels in FL are replaced by the experience records, states, and actions in deep reinforcement learning. According to environment partitioning, federated reinforcement learning can be divided into two types \cite{b20}. In horizontal federated reinforcement learning (HFRL), different agents have similar action and state spaces, and they act independently in different environments. The purpose of federating is to accelerate the exploration of the environment by experience sharing. In vertical federated reinforcement learning (VFRL), all the agents act in the same environment. The state and action space of agents can be heterogeneous, and their actions collaboratively change the environment and influence the reward.

\subsection{Proposed VFRL Coordination Algorithm}

\begin{figure}[t]
\centerline{\includegraphics[width=3.0in]{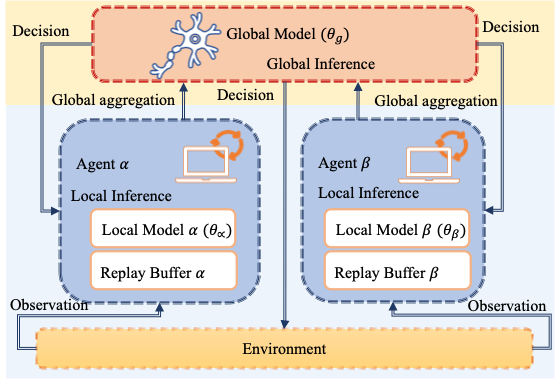}}
\caption{Process of federated deep reinforcement learning.}
\label{fig2}
\vspace{-15pt}
\end{figure}

Fig.~\ref{fig2} shows the structure of our proposed VFRL algorithm. The general process of one update cycle in VFRL can be described as follows. Firstly, each agent observes their states from the environment. Next, the local models of different agents are respectively updated with local training. The intermediate results calculated by local models are then submitted to a global model for global training, and the decisions of the global model are sent back to agents and executed to change the environment. Local models are then updated based on the feedback of the global model and the reward from the environment.

In Fig.~\ref{fig2}, agent $\alpha$ and agent $\beta$ refer to the power control xAPP and radio resource allocation xAPP, respectively. $\theta_{g}$, $\theta_{\alpha}$ and $\theta_{\beta}$ respectively denote the parameters of the global model and the local models of two xAPPs. The core idea of our proposed algorithm is to use the local Q-table generated by two xAPPs to indicate the expected reward of independent actions and make a global calibration of the local models.

The federating process can be summarized as three steps:

\begin{itemize}
    \item Local inference. During the local inference step, power control and radio resource allocation xAPPs observe state information from BSs, then they use observed state and local DQN model to calculate the local Q-table, which can be formulated as:
    \begin{equation}
    \boldsymbol{Q_{\alpha}} = DQN^{\alpha}(s_{\alpha}, \boldsymbol{a_{\alpha}} ;\theta_{\alpha})\label{eq16}
    \end{equation}
    \begin{equation}
    \boldsymbol{Q_{\beta}} = DQN^{\beta}(s_{\beta}, \boldsymbol{a_{\beta}} ;\theta_{\beta})
    \label{eq17}
    \end{equation}
    where $DQN^{\alpha}$ and $DQN^{\beta}$ denote two local DQN models of agent $\alpha$ and agent $\beta$. $s_{\alpha}$ and $s_{\beta}$ denote the locally observed states and $\theta_{\alpha}$ and $\theta_{\beta}$ denote the parameters of two DQN models.
    \item Global aggregation. During the global inference step, the two local Q-tables generated by global power control and radio resource allocation xAPPs are submitted and combined as the input of the global model to calculate a joint global Q-table, which can be formulated as:
    \begin{equation}
    \boldsymbol{Q_{g}} = [\boldsymbol{Q_{\alpha}^g}|\boldsymbol{Q_{\beta}^g}] =  DNN^{g}([\boldsymbol{Q_{\alpha}}|\boldsymbol{Q_{\beta}}] ;\theta_{g})
    \label{eq18}
    \end{equation}
    where $Q_{\alpha}^g$ denotes the Q-table of agent $\alpha$ after global calibration and $[\boldsymbol{Q_{\alpha}^g}|\boldsymbol{Q_{\beta}^g}]$ denotes the global Q-table is a splicing of Q-tables of two agents after global calibration.
    \item Global inference. During the global inference step, the global Q-table is split into two calibrated Q-tables and the split Q-tables are used for action selection, which can be formulated as:
    \begin{equation}
    a_{\alpha} = \underset{a_{\alpha}}{argmax}\boldsymbol{Q_{\alpha}^g}
    \label{eq19}
    \end{equation}    
    \begin{equation}
    a_{\beta} = \underset{a_{\beta}}{argmax}\boldsymbol{Q_{\beta}^g}
    \label{eq20}
    \end{equation}
    The actions decided by global inference will be executed in the environment and the corresponding rewards will be stored in the replay buffers. In addition, an adaptive $\epsilon$-greedy learning strategy is adopted, which means the agent will randomly choose actions with a certain probability in order to explore the unknown action space.
\end{itemize}

\begin{table}[t]
\caption{Simulation settings.}
\label{table1}
\renewcommand{\arraystretch}{1.2}
\begin{tabular}{ll}
\hline
Parameters                                                        & Settings    \\ \hline
Carrier configuration                                             & \begin{tabular}[c]{@{}l@{}}Bandwidth: 20MHz, \\ Number of RBs: 100, \\ Subcarriers of each RB: 12,\\ Maximum transmission power: 30 dBm\\ Tx/Rx antenna gain: 15 dB\\ TTI size: 2 OFDM symbols\end{tabular} \\ \hline
Propagation model                                                 & \begin{tabular}[c]{@{}l@{}}128.1 + 37.6log(Distance), \\ Log-Normal shadowing 8 dB.\end{tabular}                                                                                                            \\ \hline
Retransmissions                                                   & \begin{tabular}[c]{@{}l@{}}Maximum number of retransmissions: 1,\\ Round trip delay: 4 TTIs\end{tabular}                                                                                                    \\ \hline
Traffic mode                                                      & \begin{tabular}[c]{@{}l@{}}eMBB: Constant bit rate traffic, \\Packet size 32 Bytes, \\ URLLC: Poisson traffic, \\Packet size 16 Bytes\end{tabular}                                                                                                          \\ \hline
\begin{tabular}[c]{@{}l@{}}Networking \\ environment\end{tabular} & \begin{tabular}[c]{@{}l@{}}2 gNBs, 12 users for each BS\\ Inter-gNB distance: 500m\end{tabular}                                                                                                   \\ \hline
Slice settings                                                    & \begin{tabular}[c]{@{}l@{}}2 eMBB slices and 2 URLLC slices\\  in each BS,\\ Priority:\\ URLLC 2 \textgreater URLLC 1  \textgreater eMBB 2\textgreater eMBB 1\end{tabular}                        \\ \hline
\end{tabular}
\vspace{-15pt}
\end{table}

\section{Numeric results}
\label{s6}
\subsection{Simulation settings}
In the simulations, we consider 2 BSs, and each BS has 4 slices, 2 eMBB slices and 2 URLLC slices. Each slice has 3 users. We use constant bit rate traffic for eMBB slices while the URLLC traffic follows a Poisson distribution. Simulations are implemented with Matlab 5G toolbox and run 10 simulations with 5000 TTIs. Table \ref{table1} shows the settings of our simulations. 

We compare three cases, namely independent reinforcement learning (IRL), centralized reinforcement learning (CRL), and federated reinforcement learning (FRL). IRL means the local models of agents are trained separately without the global model. CRL means putting two xAPPs into a single agent and training a single model with joint states and actions. Here we use CRL as an optimal baseline, since the centralized control with global vision will generally bring the best performance. FRL is our proposed algorithm, in which local models are federated by a coordination model. During the simulations, We fix the URLLC traffic, change the traffic load of eMBB slices.

\subsection{Simulation results}

\begin{figure}[t]
\centerline{\includegraphics[width=7.9cm,height=5cm]{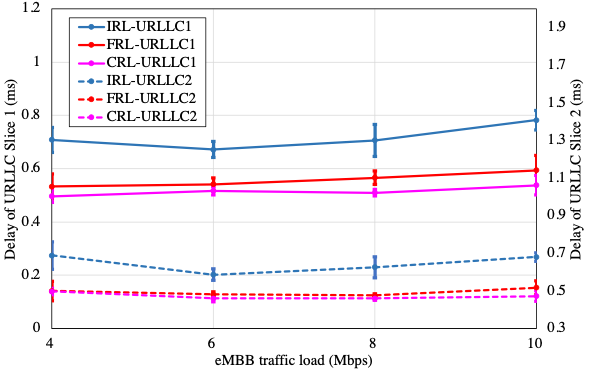}}
\caption{URLLC slices delay under various eMBB traffic load.}
\label{fig5}
\vspace{-5pt}
\end{figure}

\begin{figure}[t]
\centerline{\includegraphics[width=7cm,height=5cm]{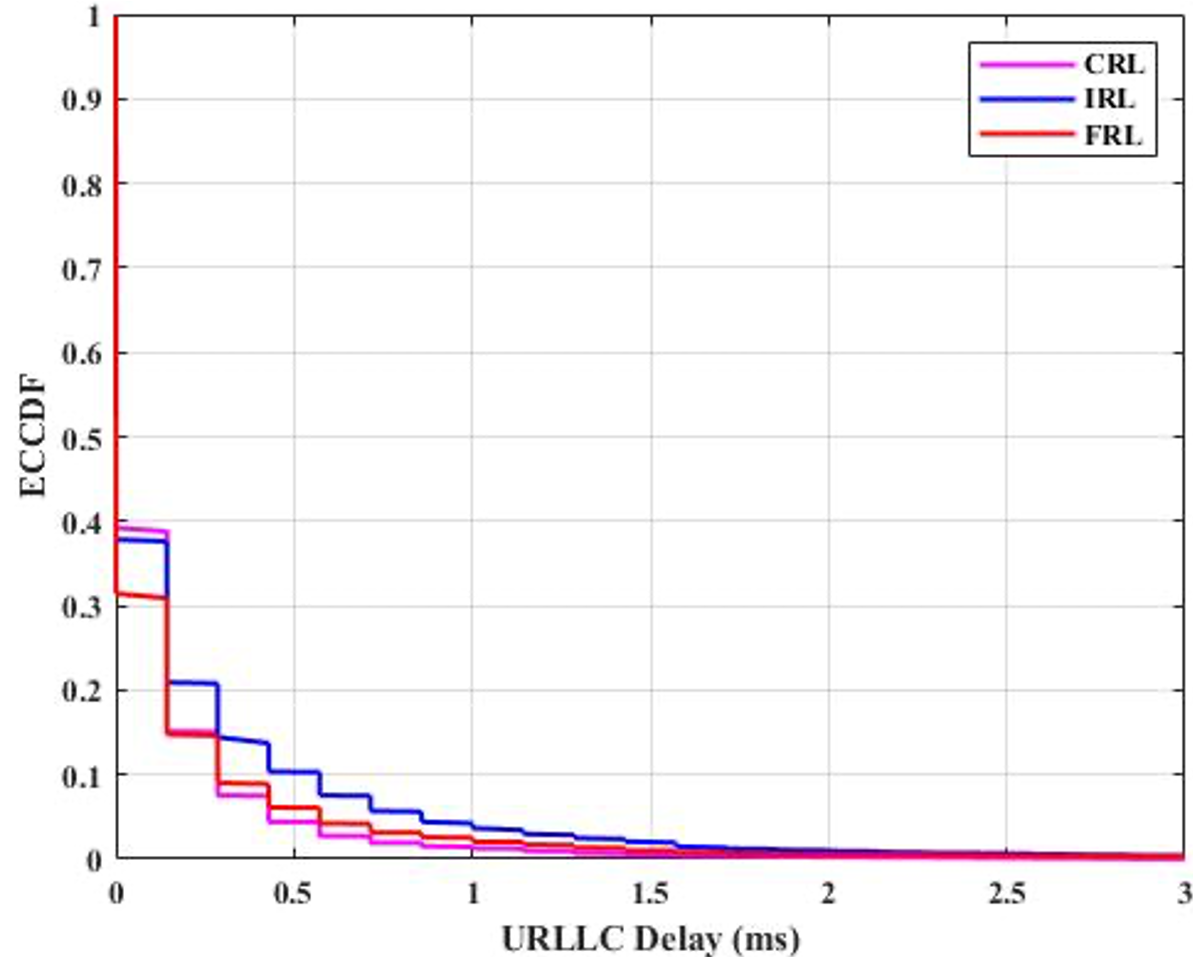}}
\caption{ECCDF of the URLLC slices delay distributions.}
\label{fig6}
\vspace{-15pt}
\end{figure}

\begin{figure}[t]
\centerline{\includegraphics[width=2.9in]{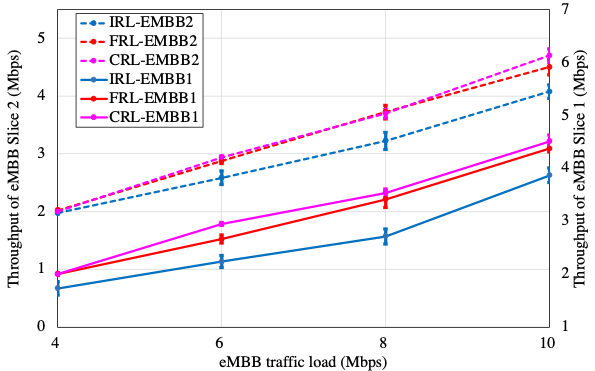}}
\caption{Throughput of eMBB slice under various eMBB traffic load.}
\label{fig7}
\vspace{-15pt}
\end{figure}

Fig. ~\ref{fig5} shows the delay of URLLC slices when the URLLC traffic load is 2 Mbps and the eMBB traffic load changes from 4 Mbps to 10 Mbps. Since the priority of URLLC slices is higher than that of eMBB slices and the traffic load of URLLC slices is constant, the delay of URLLC slices does not significantly increase when the eMBB traffic load changes. It can be observed that the delay of URLLC slices with the FRL algorithm is lower than that with the IRL algorithm and is very close to that with CRL. It means the federating process can help coordinate different xAPPs and achieve comparable network performance with centralized training. Specifically, when the eMBB traffic load is 10 Mbps, the delay with FRL is 33\% lower than the IRL algorithm. Fig.~\ref{fig6} further shows the empirical complementary cumulative distribution function (ECCDF) of the delay of URLLC slices when the eMBB traffic load is 8 Mbps. It shows that our proposed FRL has a better delay distribution than IRL, which is indicated by a lower ECCDF curve.  

Moreover, Fig.~\ref{fig7} shows the throughput of eMBB slices at various eMBB traffic loads. When the eMBB traffic load grows, more eMBB traffic needs to be transmitted, so the throughput of eMBB slices will also grow. We can observe that FRL can achieve higher eMBB throughput compared with IRL cases, and the network performance of FRL is very close to IRL. Specifically, when the eMBB traffic load is 10 Mbps, the eMBB throughput of FRL is 11\% higher than that of IRL.

\begin{figure}[t]
\centerline{\includegraphics[width=2.8in]{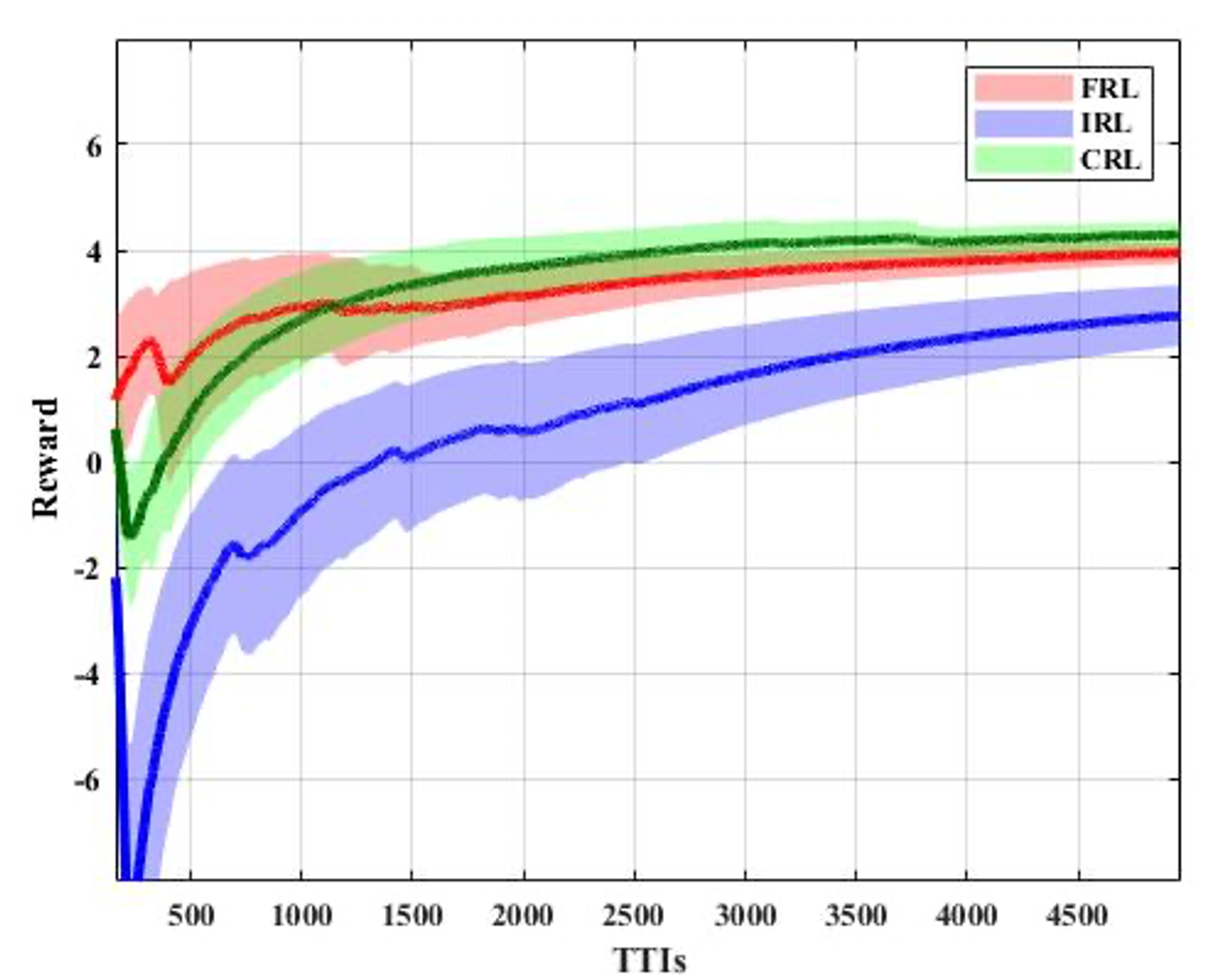}}
\caption{Convergence comparison.}
\label{fig4}
\vspace{-20pt}
\end{figure}

Finally, the Fig.~\ref{fig4} shows the convergence performance of IRL, CRL, and FRL when the eMBB traffic is 8 Mbps, and the URLLC traffic is 2 Mbps. We can observe that CRL can ultimately achieve the highest reward, while the converged reward of FRL is slightly lower than CRL. This is because CRL uses a central training approach, which allows for more comprehensive global information to be captured. However, the FRL has a faster convergence speed than CRL, which can be explained by the huge state and action space of CRL approach. The reward of IRL is much lower than the other two algorithms. Based on the convergence trend, FRL converges faster than CRL.

We can draw the conclusion that the proposed FRL algorithm can achieve better network performance than IRL, and the performance of FRL is very comparable with the optimal baseline CRL. Meanwhile, FRL enables faster convergence and can scale to multiple agent situations with distributed structure. It also allows information observed by different agents to be kept confidential and consumes fewer training resources.

\section{Conclusion}
\label{s7}
Federated deep reinforcement learning has emerged as one of the most state-of-the-art machine learning techniques. Here, we proposed a federated deep reinforcement learning solution to coordinate two xAPPs, power control, and radio resource allocation for network slicing in O-RAN. The core idea is to make distributed agents submit their local Q-tables to a coordination model for federating, and generate a global Q-table for action selection. According to the simulation results, our proposed algorithm can obtain lower delay and higher throughput compared with independent reinforcement learning cases, and the network performance is comparable to the centralized training results. The proposed framework can be scaled up to more than two xAPPs and the increased computational cost is equal to the computational cost required to train a neural network whose input and output dimensions are the sum of action space dimensions of all the xAPPs.
\vspace{-3pt}
\section*{Acknowledgement}
This work is funded by Canada Research Chairs program and NSERC Collaborative Research and Training Experience Program (CREATE) under Grant 497981.

\end{document}